\documentclass[final]{svjour3}
\usepackage{graphicx}
\usepackage{rotating}
\usepackage{amssymb}
\usepackage{mathptmx}
\usepackage{multirow}
\usepackage[square, super, sort, numbers]{natbib}
\usepackage{siunitx}
\usepackage[caption=false]{subfig}
\usepackage{threeparttable}
\makeatletter
\journalname{Journal of Low Temperature Physics}

\bibpunct{}{}{,}{s}{}{,}

\begin{document}

\newcommand{\HeIII}{He\textsuperscript{3}}
\newcommand{\hdblarrow}{H\makebox[0.9ex][l]{$\downdownarrows$}-}
\title{A Continuous 100-mK Helium-Light Cooling System for MUSCAT on the LMT}

\author{T.~L.~R.~Brien \and
E.~Castillo-Dominguez \and
S.~Chase \and
S.~M.~Doyle}
\institute{tom.brien@astro.cf.ac.uk \\[1.5EX]
\small{
T. L. R. Brien and S. M. Doyle: 
\at School of Physics and Astronomy, Cardiff University, Cardiff, CF24 3AA, United Kingdom
\and 
E.~Castillo-Dominguez: 
\at Instituto Nacional de Astrof\'{i}sica, \'{O}ptica y Electr\'{o}nica, Luis Enrique Erro \#1, Tonantzintla, Pue, M\'{e}xico, ZIP 72840
\and
S. Chase: 
\at Chase Research Cryogenics Ltd, Uplands, 140 Manchester Road, Sheffield, S10 5DL, United Kingdom
}}

\maketitle

\begin{abstract}
The MUSCAT instrument is a large-format camera planned for installation on the Large Millimeter Telescope (LMT) in 2018. MUSCAT requires continuous cooling of several large-volume stages to sub-Kelvin temperatures, with the focal plane cooled to $100~\si{\milli\kelvin}$. Through the use of continuous sorption coolers and a miniature dilution refrigerator, the MUSCAT project can fulfil its cryogenic requirements at a fraction of the cost and space required for conventional dilution systems. Our design is a helium-light system, using a total of only 9 litres of helium-3 across several continuous cooling systems, cooling from $4~\si{\kelvin}$ to $100~\si{\milli\kelvin}$. Here we describe the operation of both the continuous sorption and the miniature dilution refrigerator systems used in this system, along with the overall thermal design and budgeting of MUSCAT. MUSCAT will represent the first deployment of these new technologies in a science-grade instrument and will prove the concept as a viable option for future large-scale experiments such as CMB-S4.
\end{abstract}

\section{Introduction}
The Mexico UK Sub-mm Camera for AsTronomy (MUSCAT) will be a $1\mbox{-}\si{\milli\metre}$ receiver consisting of 1,800 lumped-element kinetic inductance detectors (LEKIDs) operating at the photon noise limit. MUSCAT is scheduled to be installed on the Large Millimeter Telescope (LMT) (Puebla, Mexico) in 2018 following the upgrade of the LMT's primary mirror from $32~\si{\metre}$ to $50~\si{\metre}$. MUSCAT will be one of the first new instruments to be installed on the 50-m LMT. To enable optimum operation of MUSCAT, a cryogenic system has been designed capable of continuously cooling the focal plane of MUSCAT to a temperature of approximately $100~\si{\milli\kelvin}$. Additionally, this system will cool the internal structure of the MUSCAT cryostat and provide appropriate heat-sinking and radiation-shielding stages. While cooling of detectors to $100~\si{\milli\kelvin}$ is desired for optimum instrument performance---along with future proofing---the key target for the system is to reduce the thermal load on the detector stage such that it is dominated by the expected sky load of $\sim 1.65~\si{\micro\watt}$ while cooling this stage to below $200~\si{\milli\kelvin}$; the interfaces and cooling systems have been selected with consideration to this objective. However, access to these lower temperatures may prove beneficial for future upgrades requiring higher detector sensitivities.
\section{Cooling Chain}
\begin{figure}[tbh]
\centering
\subfloat[]{
\includegraphics[height=0.25\textheight]{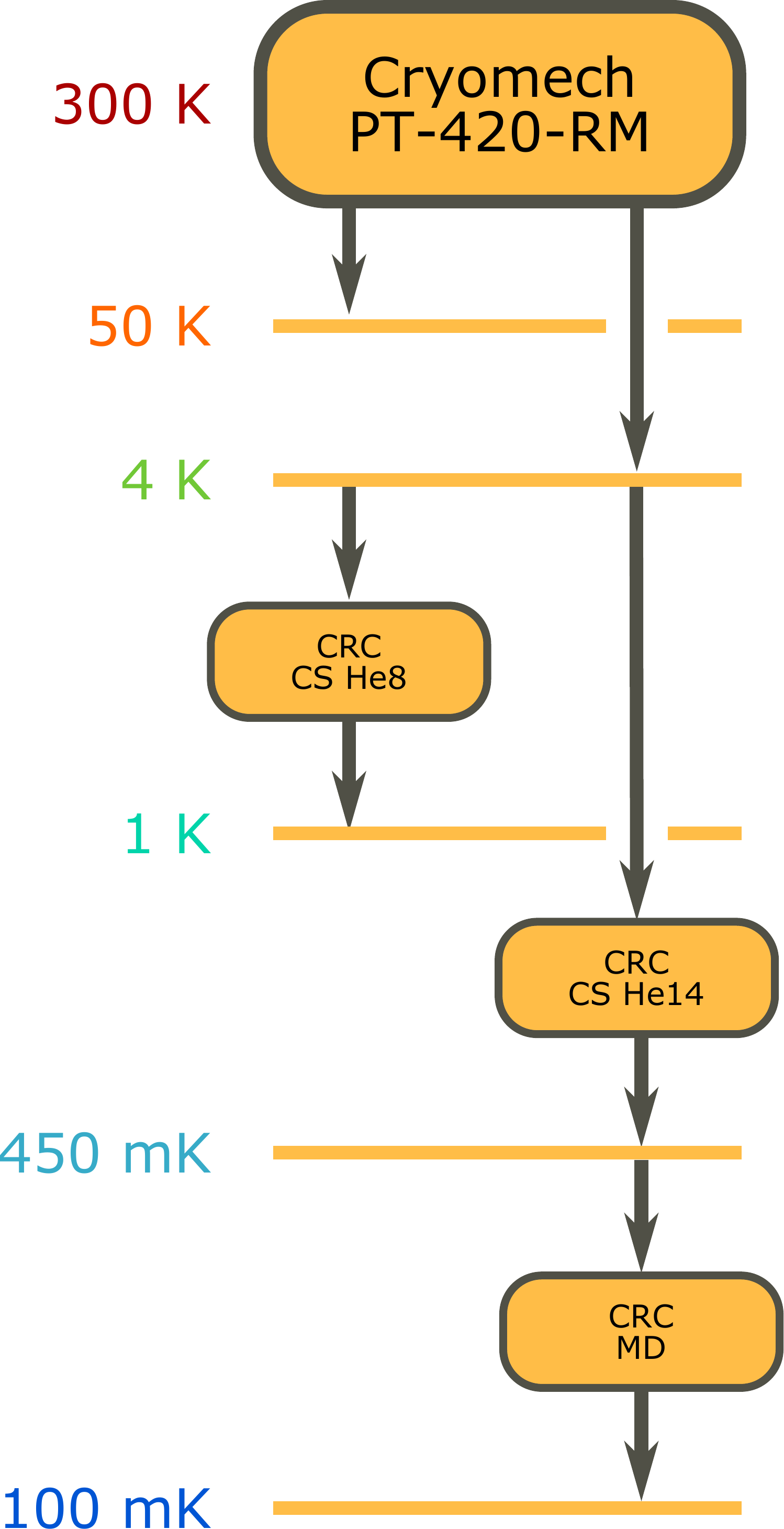}
\label{fig:coolingChain} }
\subfloat[]{\includegraphics[width=0.7\textwidth]{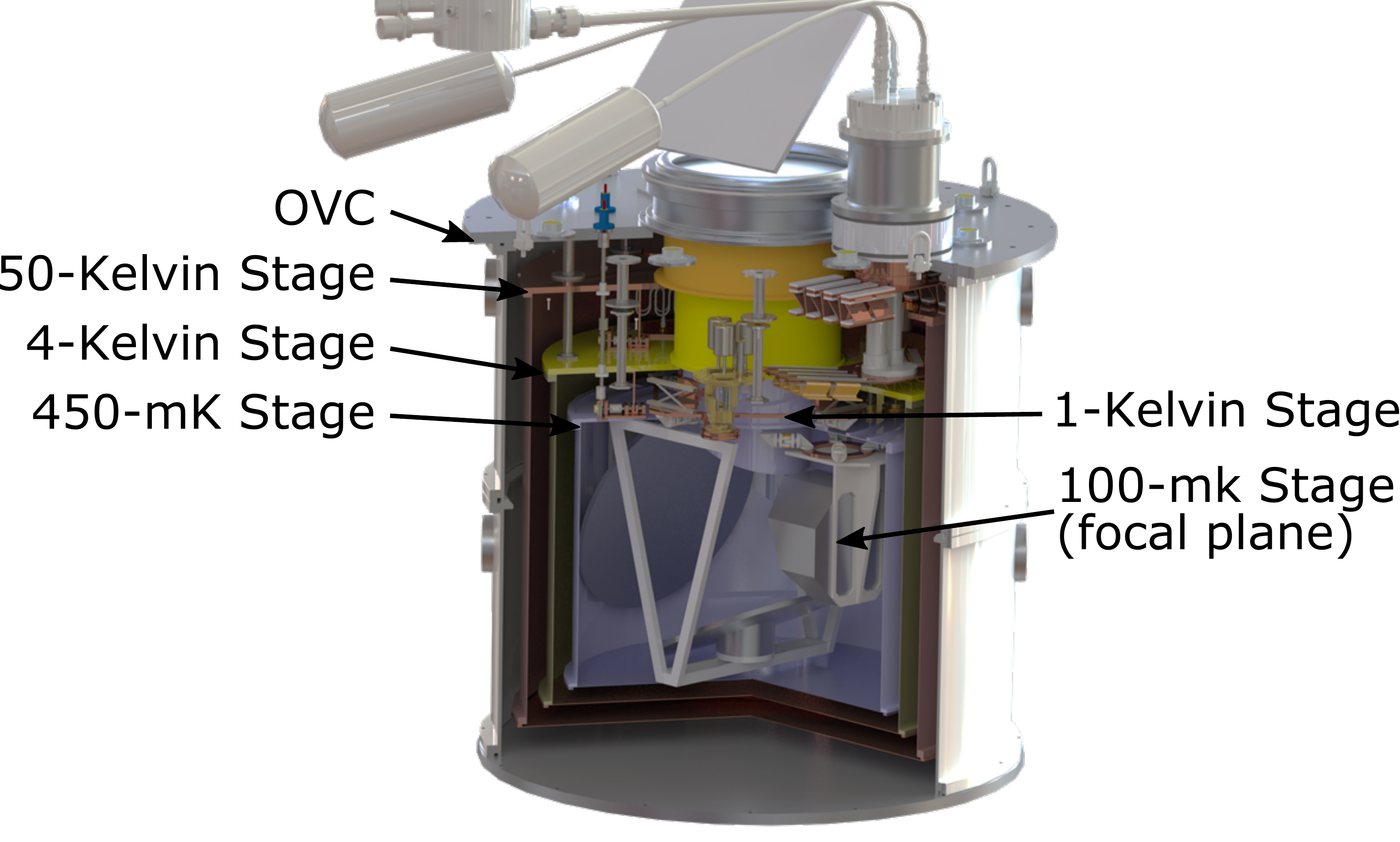}
\label{fig:xSection} }
\caption{(a) The MUSCAT cooling system from room temperature ($300~\si{\kelvin}$) to the focal plane ($100~\si{\milli\kelvin}$). (b) Cross-sectional view of the CAD design for the MUSCAT cryostat. The 50-Kelvin, 4-Kelvin, and 450-mK stages are coloured for illustration only. The two continuous sorption coolers are shown to greater detail in Fig.~\ref{fig:SorpPumps}. (Colour figure online.)}
\end{figure}
The MUSCAT cooling chain contains a total of four separate cooling systems, namely: a Cryomech Inc.\footnote{Cryomech Inc., 13 Falso Drive Syracuse, New York 13211, USA} PT-420-RM Pulse Tube Cooler (PTC), two Chase Research Cryogenics Ltd\footnote{Chase Research Cryogenics Ltd, Uplands, 140 Manchester Road, Sheffield, S10 5DL, UK} (CRC) continuous sorption (CS) coolers, and a final Chase Research Cryogenics Ltd miniature dilution (MD) refrigerator.
\par 
The pulse tube cooler is used to cool two stages of the cryostat; the first of these operates at a nominal temperature of $50~\si{\kelvin}$ and is used as a heat sinking point and to cool a radiation shield; the second stage is cooled to a nominal temperature of $4~\si{\kelvin}$ and contains an additional radiation shield, readout amplifiers and is also used as the thermal bath for the two continuous sorption coolers. The two continuous sorption coolers cool separate stages down to temperatures of $1~\si{\kelvin}$ and $450~\si{\milli\kelvin}$, both of these stages fluctuate in temperature during operation (see Fig.~\ref{fig:flipFlop_cycle}) however modelling of this behaviour along with preliminary testing has shown that these fluctuations should not be an issue for MUSCAT. The 1-Kelvin stage provides heat sinking for readout and housekeeping cabling prior to the final stages whereas the 450-mK stage contains the final radiation shield and acts as the condensing point for the miniature dilution refrigerator. The 450-mK stage's radiation shield defines the optical enclosure for the detector array. The MUSCAT cooling chain is illustrated in Figure~\ref{fig:coolingChain}. The full mechanical design of the MUSCAT cryostat is described by Castillo-Dominguez et al.\cite{Castillo2017} but is outlined in Figure~\ref{fig:xSection}.
%
%
\section{The Continuous Sorption Coolers}
\begin{figure}[htb]
\centering
\subfloat[]{
\includegraphics[scale = 0.25]{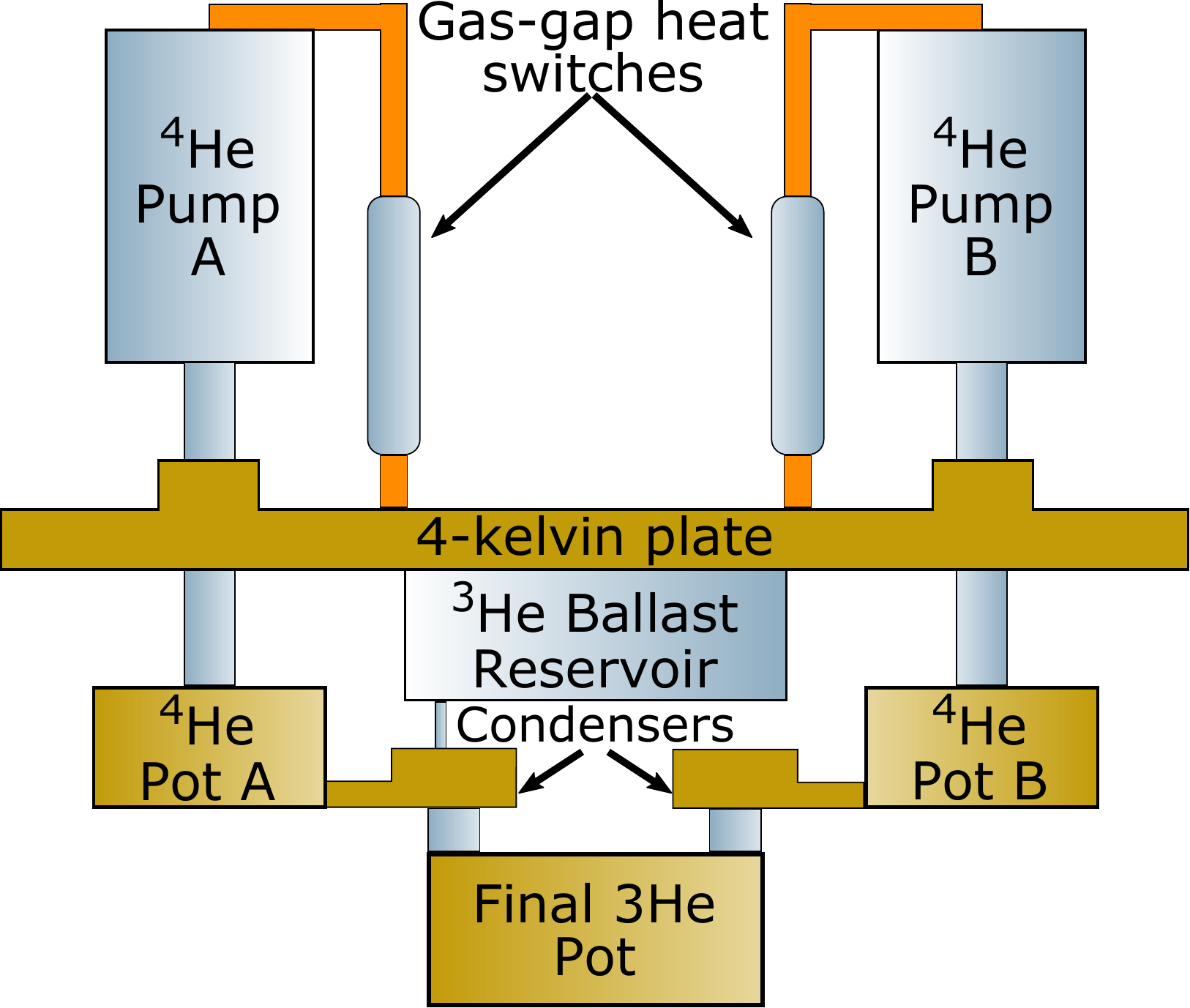}
\label{fig:He8Pump}}
\subfloat[]{
\includegraphics[scale = 0.25]{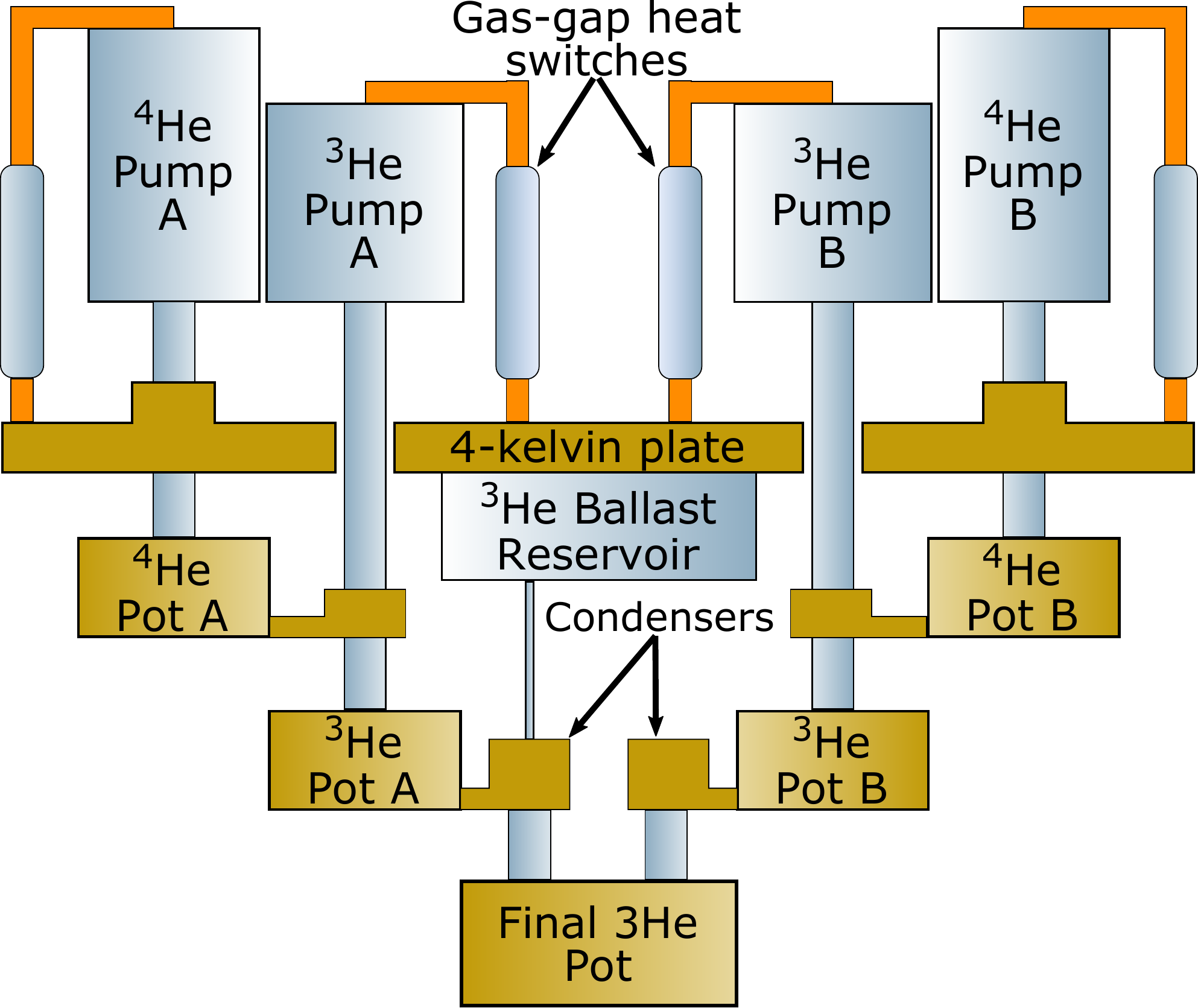}
\label{fig:He14Pump}}
\caption{Schematics of the two continuous sorption coolers used in MUSCAT ((a) 1-Kelvin cooler, (b) 450-mK cooler). The \HeIII{} ballast is maintained as a liquid in the final head since at any given time one of the two condensers is at its base temperature---this will be below the temperature achieved in the final head. This ballast ensures there is sufficient thermal mass to keep the final head cold while one of the sub systems is regenerated. Both systems operate on the same principle described by Klemencic et al.\cite{klemencic2016} (Colour figure online.)}
\label{fig:SorpPumps}
\end{figure}
The third and fourth temperature stages of MUSCAT (the first two not cooled by the pulse tube cooler) are cooled by continuous sorption (CS) coolers. Each of these systems operate essentially as two separate subsystem cycled in anti phase with each other. Both subsystems cool the same final evaporator which contains a small amount of a helium-3 ballast gas. The first of these systems contains two helium-4 pumps and is designed to achieve a minimum an operating temperature of approximately $1~\si{\kelvin}$.
\par
\begin{figure}[hbt]
\begin{center}
\includegraphics[width=0.8\textwidth]{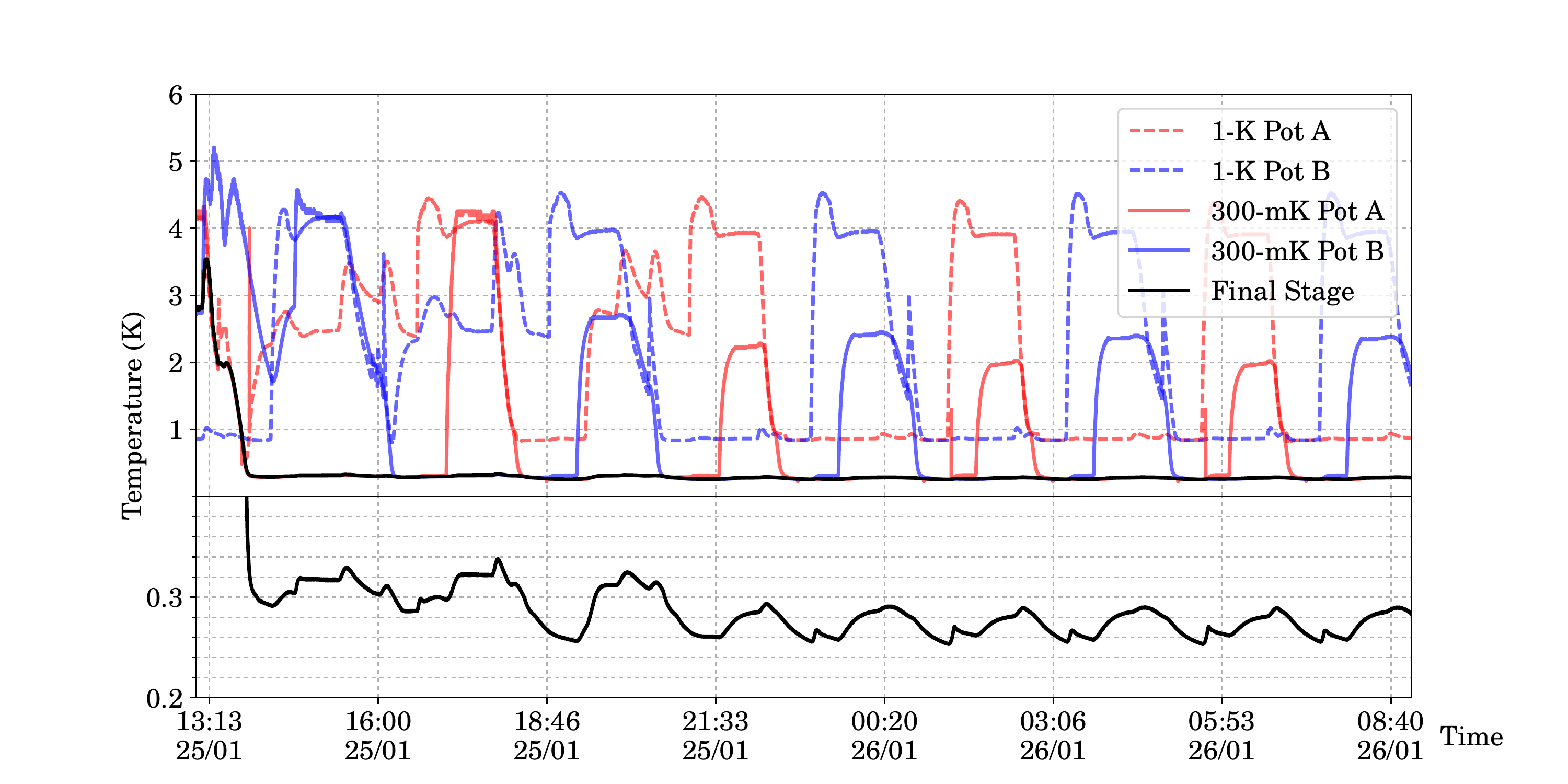}
\end{center}
\caption{Initial cycles of the two systems (red and blue lines) of the second MUSCAT continuous sorption cooler, used to cool the 450-mK stage. For these tests the system was under minimal loading. The final stage temperature is shown by the black line. When the coloured lines are high, that system is being charged with gas, when the lines are low that system is being pumped upon and is cooling the final evaporator. (Colour figure online.)}
\label{fig:flipFlop_cycle} 
\end{figure}

The second CS system, used to cool the 450-mK stage of MUSCAT, consists of two pairs of helium-4 and helium-3 pumps. The performance of a similar system has been described previously and has been shown to operate at a temperature of $300~\si{\milli\kelvin}$ under minimal thermal loading and to be thermally stable for in excess of three months.\cite{klemencic2016} Our system is expected to operate at a higher temperature due to the presence of a higher thermal load, as described later. The temperature at the final head of the CS cooler fluctuates naturally during its operation due to the constant recycling of the system. These fluctuations can be seen in the lower panel of Figure~\ref{fig:flipFlop_cycle}, which shows the first five cycles of each subsystem over the a period of approximately 20 hours under minimal loading. The black line in Figure~\ref{fig:flipFlop_cycle} (shown zoomed in in the lower panel) shows the temperature of the final evaporator of this system, it is seen that for the first two cycles of each subsystem the minimum temperature achieved decreases for each cycle but the final performance is not achieved until the third cycle of the A (red) system (the red or blue traces in the upper panel are low when that system is being pumped upon and is thus providing cooling). In the \textit{steady state} (after the first two cycles of each subsystem) the temperature of the final stage fluctuates between $255\mbox{--}290~\si{\milli\kelvin}$. It has been shown that through PID control of a heater mounted to this head, the achieved temperature can be stabilised at $365.0\pm0.1~\si{\milli\kelvin}$ in the presence of a $20\mbox{-}\si{\micro\watt}$ thermal load (designed to represented the device load for a particular application).\cite{klemencic2016} PID control of these units is not required for MUSCAT as the final cooler provides sufficient thermal dampening to nullify the oscillations at the detector.
\par 
\begin{figure}[hbt]
\centering
\subfloat[]{
\includegraphics[width = 0.48\textwidth]{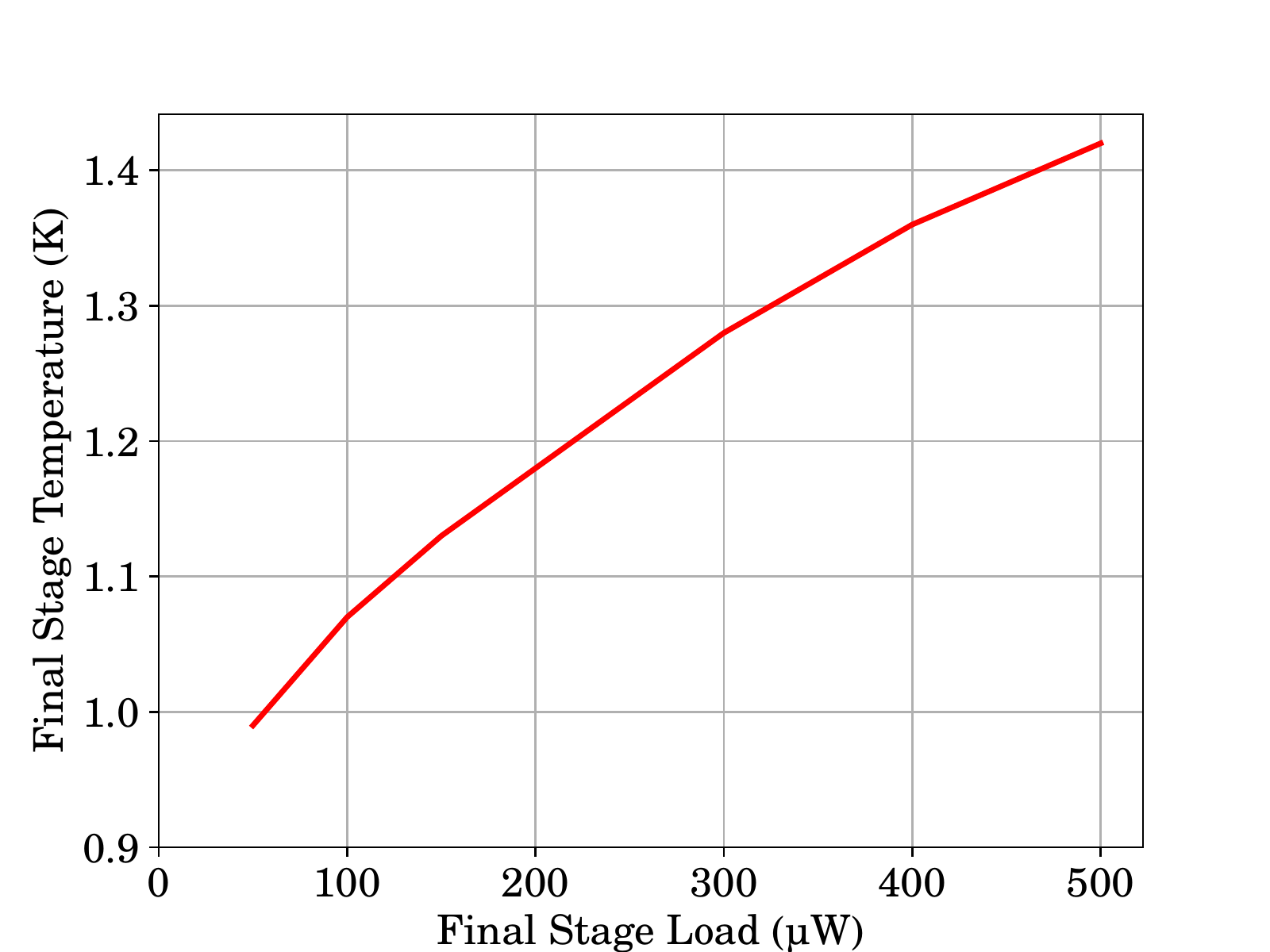}
\label{fig:1Kload}}
\subfloat[]{
\includegraphics[width = 0.48\textwidth]{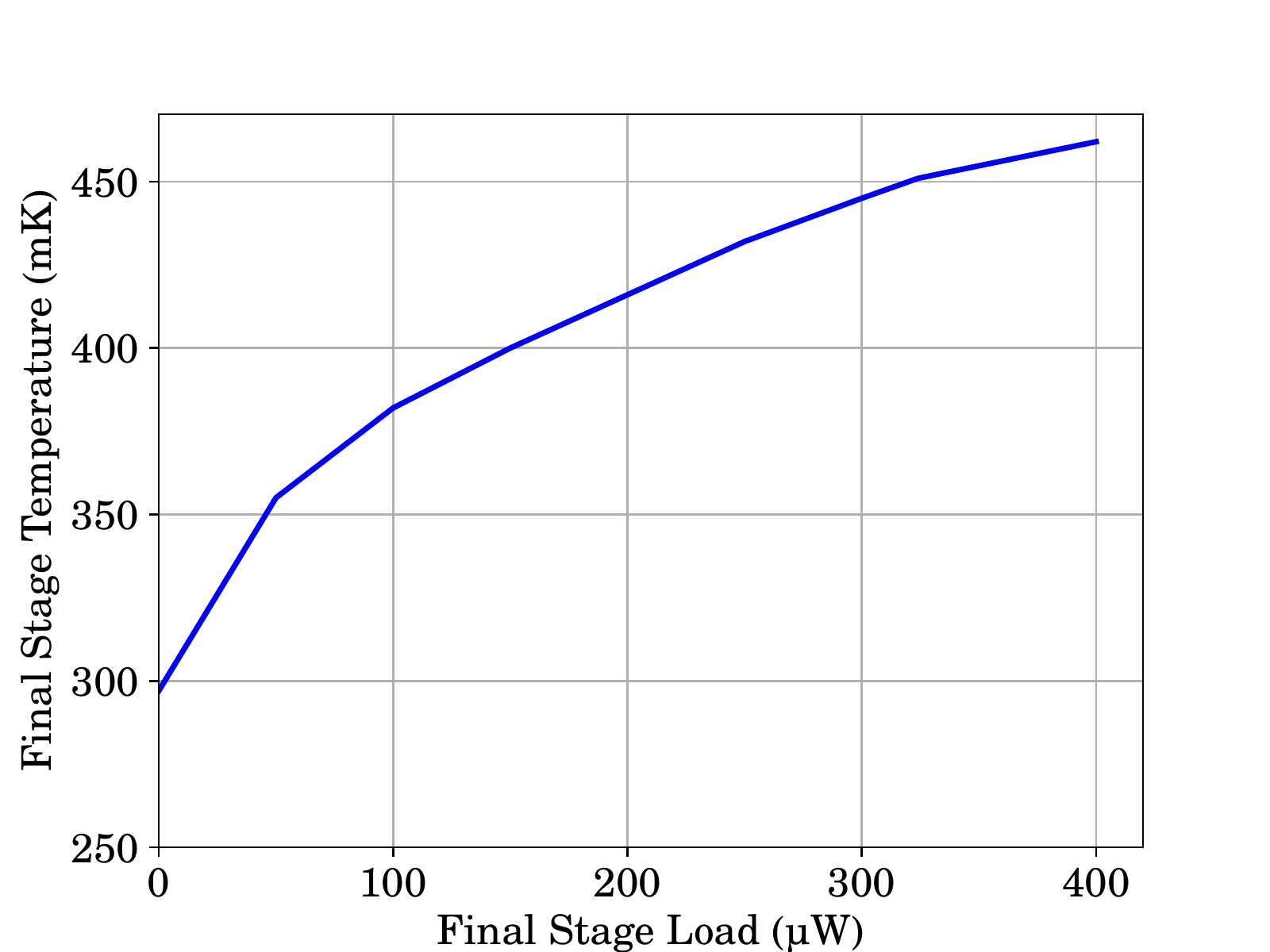}
\label{fig:450mKload}}
\caption{Measured load curves for the two continuous sorption coolers used by MUSCAT. (a) 1-Kelvin cooler; (b) 450-mK cooler. (Colour figure online.)}
\label{fig:CSloadCurves}
\end{figure}
In order to operate at the levels of thermal loading anticipated for MUSCAT, the timing and pump temperatures used to cycle each subsystem have been tuned to ensure that the system being regenerated is fully prepared prior to the liquid \HeIII{} being depleted in the other system. Using this scheme a load curve for a prototype of each of the two CS coolers has been measured. These load curves are shown in Figure~\ref{fig:CSloadCurves} and the stated temperatures correspond to the maximum temperature observed during the steady-state fluctuations.

\section{The Miniature Dilution Refrigerator}
The final cooling stage of MUSCAT will be a miniature dilution (MD) refrigerator. This system works through the same principle of moving helium-3 molecules across the phase boundary between a \HeIII-rich phase and a \HeIII-poor phase as a conventional dilution refrigerator.\cite{London1962,Das1965} Unlike the conventional dilution refrigerator, our system is fully contained within the structure of the cryostat and requires no mechanical pumps or large gas-handling systems. Instead the flow of \HeIII\ is driven by a temperature difference between the evaporator (also referred to as the \textit{still}) of the dilution refrigerator and a condensing point. So long as the condenser is cooler than the still, \HeIII\ molecules from the \HeIII-poor phase evaporated in the still will condense in the dilutor's condenser and are then returned as liquid to the mixing chamber, under gravity, where they join the \HeIII-rich phase. A key advantage of this architecture is that the helium-3 requirement for a miniature dilution refrigerator is of order only $2~\si{\liter}$ and requires no active pump. The achieved minimum temperature of such a unit is limited by the circulation rate of \HeIII{} molecules (as for any dilution refrigerator).
\par
MUSCAT will utilise a second-generation miniature dilution refrigerator from Chase Cryogenics Ltd (CRC), this unit is currently undergoing the finally stages of commissioning. However, a first-generation CRC miniature dilution refrigerator has been run in a general-purpose testing cryostat at Cardiff since 2014 and achieves a minimum temperature of $77~\si{\milli\kelvin}$ under minimal load and a sustained temperature of $88~\si{\milli\kelvin}$ under a $5\mbox{-}\si{\micro\watt}$ thermal load. The base temperature of the dilutor is limited by the slower circulation rate in this pump-free system and as such is higher the a pump driven system. This unit has been selected in part due to the compact size of this cooler and its relatively simple operation (requiring only heating of the still). Furthermore, the use of this technology aids MUSCAT's secondary objective of acting as an on-sky technology demonstrator not only for such a cooling technique but also access to temperatures of order $100~\si{\milli\kelvin}$ may prove vital in future upgrades of MUSCAT. For example, should an on-chip spectrometer (an option being strongly considered as a potential upgrade) be deployed on MUSCAT, the optical power on each detector may be of order 1,000 times less than when operating in the continuous 1.1-mm band thus potentially requiring a lower operating temperature to decrease the device limit of the noise-equivalent power, enabling higher sensitivities.
\section{Thermal Supports}
\begin{figure}[htb]
\centering
\subfloat[]{
\includegraphics[width = 0.35\columnwidth]{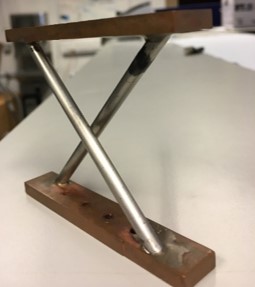}
\label{fig:crossBeam}}
\hspace{0.1\columnwidth}
\subfloat[]{
\includegraphics[width = 0.35\columnwidth]{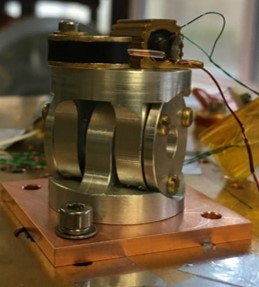}
\label{fig:SapphireJoint}}
\caption{Mechanical supports selected for use between the final stages of MUSCAT. (a) thin-walled stainless steel cross beam, (b) crushed-sapphire joint (shown with thermometer and heater installed for characterisation). (Colour figure online.)}
\label{fig:supports}
\end{figure}

Prior to modelling the thermal behaviour of MUSCAT, it is necessary to characterise the thermal supports used between the coldest stages of the cryostat. Two supports have been identified for use at these stages of MUSCAT. These are a thin-walled stainless steel crossbeam and a crushed-sapphire powder joint as developed for the SCUBA-2 instrument.\cite{Bintley2007} These supports are shown in Figure~\ref{fig:supports}.
\par
\begin{figure}[tb]
\begin{center}
\includegraphics[width=0.6\textwidth]{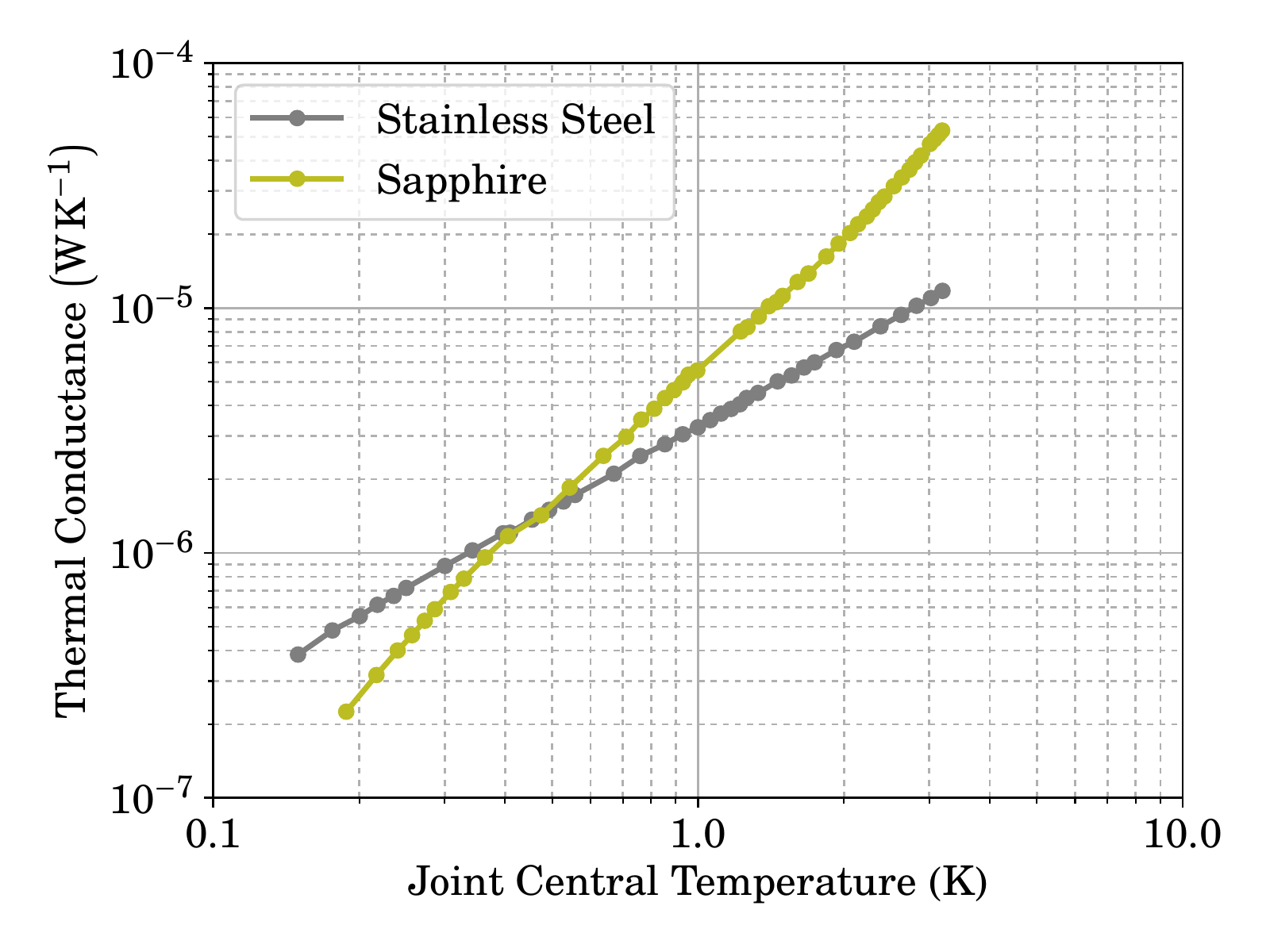}
\end{center}
\caption{Comparison of the thermal conductance of a thin-walled stainless steel crossbeam (grey) and crushed-sapphire joint (yellow). It is clear that for configurations where the central temperature of the joint will be less than $450~\si{\milli\kelvin}$ the use of the crushed-sapphire joint is preferable. (Colour figure online.)}
\label{fig:supportThermalCond} 
\end{figure}

 
These supports---particularly the crushed-sapphire powder joints---represent complex mechanical assemblies with multiple interfaces, as such modelling of their thermal performance is both complex and inaccurate, as such we have opted instead to measure their thermal conductance of the joints directly at the relevant temperatures. In order characterise the performance of these supports in the sub-Kelvin range, one end of each support was mounted to the 100-mK plate of a cryostat cooled by a miniature dilution refrigerator, a temperature-stable heater and germanium thermometer were mounted on the other end of the support (the \textit{isolated} end, as seen in Figure~\ref{fig:SapphireJoint}) to apply a known thermal load and measure the temperature difference across the joints. Measurements were taken for three heat-sink temperatures of $100~\si{\milli\kelvin}$, $350~\si{\milli\kelvin}$, and $1.2~\si{\kelvin}$ corresponding to the nominal temperatures of the three MUSCAT stages of interest. At each bath temperature, numerous powers were applied through a load heater and these data have been combined to plot the thermal conductance of the two supports, as shown in Figure~\ref{fig:supportThermalCond}.
\section{Thermal Model}
To verify the plausibility of the cooling chain presented here, we have modelled the expected thermal load on each stage of MUSCAT, the results of this are presented in Table~\ref{tab:model}. A complete description of the structure and interfaces between stages is given by Castillo-Dominguez et al.\cite{Castillo2017}
\begin{table}[tb]
\caption{Summary of thermal modelling for MUSCAT}
\label{tab:model}
\centering
\begin{threeparttable}
\begin{tabular}{llllll}\hline
 &\multicolumn{5}{c}{\textbf{Stage}}\\ 
\textbf{Consideration} & \textbf{50 K} & \textbf{4 K} & \textbf{1 K} & \textbf{450 mK} & \textbf{100 mK} \\ \hline
Mech. load through supports &$8.92~\si{\watt}$ & $64~\si{\milli\watt}$  &$116~\si{\micro\watt}$ & $12.4~\si{\micro\watt}$ & $1.04~\si{\micro\watt}$ \\ 
Rad. load from previous stage & $25.68~\si{\watt}$ & $7~\si{\milli\watt}$  & \multicolumn{1}{c}{-\tnote{{a}}} & $30.0~\si{\micro\watt}$ & $0.005~\si{\micro\watt}$ \\ 
Opt. load from filters & $3.10~\si{\watt}$ & $32~\si{\milli\watt}$  & \multicolumn{1}{c}{-\tnote{{a}}} & $22.3~\si{\micro\watt}$ & $0.019~\si{\micro\watt}$ \\ 
RF Cabling & $0.50~\si{\watt}$ & $10~\si{\milli\watt}$  &$22~\si{\micro\watt}$ & $16.1~\si{\micro\watt}$ & $0.14~\si{\micro\watt}$ \\ 
DC cabling &$0.17~\si{\watt}$ & $8~\si{\milli\watt}$  &$28~\si{\micro\watt}$ & $14.6~\si{\micro\watt}$ & $0.23~\si{\micro\watt}$ \\ 
RF amplifiers & \multicolumn{1}{c}{-} & $125~\si{\milli\watt}$ & \multicolumn{1}{c}{-}  &\multicolumn{1}{c}{-} & \multicolumn{1}{c}{-} \\ 
Cooling Systems & \multicolumn{1}{c}{-} & $0.2\mbox{--}1.2~\si{\watt}$\tnote{{b}} & \multicolumn{1}{c}{-}  & $300~\si{\micro\watt}$ & \multicolumn{1}{c}{-} \\ 
Sky load & \multicolumn{1}{c}{-} & \multicolumn{1}{c}{-} & \multicolumn{1}{c}{-} & \multicolumn{1}{c}{-} & $1.65~\si{\micro\watt}$ \\ \hline 
\textbf{Total Load} & $38.37~\si{\watt}$ & $0.5\mbox{--}1.6~\si{\watt}$  & $166~\si{\micro\watt}$ & $395.4~\si{\micro\watt}$ & $3.08~\si{\micro\watt}$ \\ 
\textbf{Expected Temperature} & $44~\si{\kelvin}$ & $2.8\mbox{--}4.1~\si{\kelvin}$ & $1.10\mbox{--}1.15~\si{\kelvin}$\tnote{{c}} & $440\mbox{--}460~\si{\milli\kelvin}$\tnote{{c}} & $<88~\si{\milli\kelvin}$ \\ \hline
\end{tabular}
\begin{tablenotes}
\item[{a}] 1-K stage has negligible surface area
\item[{b}] Load on 4-K stage fluctuates during cycling of CS coolers due to operation of heat switches
\item[{c}] Natural fluctuation of CS cooler (as seen in Figure~\ref{fig:flipFlop_cycle}) without the use of PID stabilisation
\end{tablenotes}
\end{threeparttable}
\end{table}
The total required volume of \HeIII{} in MUSCAT is approximately $9~\si{\liter_{STP}}$ of gas. This consists of 4 litres in the 450-mK cooler pumps, 1.5 litres in each of the continuous soprtion ballast tanks, and 2 litres in the miniature dilutor.
%

\section{Conclusion}
We have designed a cooling chain capable of cooling the focal plane of MUSCAT to below $100~\si{\milli\kelvin}$ continuously. The total requirement for helium-3 in the entire cooling chain is only $9~\si{\liter_{STP}}$, substantially reducing the cost of MUSCAT's cooling system compared to more conventional systems capable of comparable performance. The individual cooling systems are currently undergoing the final stages of commissioning and verifications and will be installed in the MUSCAT cryostat in the coming months.
\\[2EX]
\textbf{Acknowledgements} This work has been supported by Research Councils UK (RCUK) and Consejo Nacional de Ciencia y Tecnolog\'{i}a (CONACYT) under the Newton Fund, project ST/P002803/1. This work has also been supported by Chase Research Cryogenics Ltd.


\end{document}